\documentstyle[epsf]{article}

\newcommand{\simgt}
{\mbox{\raisebox{-0.5ex}{$\textstyle \; \sim$}
\raisebox{0.8ex}{$\textstyle \!\!\!\!\!\!\! >$}}}

\begin{document}
\mbox{ } \rightline{UCT-TP-261/03}\newline
\rightline{September 2003}\newline
\vspace{3.5cm}

\begin{center}
{\Large {\bf Finite energy chiral sum rules in QCD\footnote{{\bf Work
supported in part by the Volkswagen Foundation}}}}\\[0pt]
\vspace{0.5cm} {\bf C. A. Dominguez$^{(a)}$, K. Schilcher$^{(b)}$}\\[0.5cm]
$^{(a)}$Institute of Theoretical Physics and Astrophysics\\[0pt]
University of Cape Town, Rondebosch 7700, South Africa\\[0.5cm]
$^{(b)}$Institut f\"{u}r Physik, Johannes Gutenberg-Universit\"{a}t\\[0pt]
Staudingerweg 7, D-55099 Mainz, Germany\\[0pt]
\end{center}

\vspace{.5cm}

\begin{abstract}
\noindent

The saturation of QCD chiral sum rules of the Weinberg-type is analyzed
using ALEPH and OPAL experimental data on the difference between vector and
axial-vector correlators (V-A). The sum rules exhibit poor saturation up to
current energies below the tau-lepton mass. A remarkable improvement is
achieved by introducing integral kernels that vanish at the upper limit of
integration. The method is used to determine the value of the finite
remainder of the (V-A) correlator, and its first derivative, at zero
momentum: $\bar{\Pi}(0) = - 4 \bar{L}_{10} = 0.0257 \pm 0.0003 \;,$ 
and $\bar{\Pi}^{\prime}(0) = 0.065 \pm 0.007\; \mbox{GeV}^{-2}$. 
The dimension $d=6$ and $d=8$ 
vacuum condensates in the Operator Product Expansion are also determined:
 $<{\cal {O}}_{6}> = -(0.004 \pm 0.001) \;\mbox{GeV}^6,$ and
 $<{\cal {O}}_{8}> = -(0.001 \pm 0.006) \;\mbox{GeV}^8.$
\end{abstract}

\newpage \setlength{\baselineskip}{1.5\baselineskip} \noindent\ 
Since the pioneering work of Shifman, Vainshtein and Zakharov \cite{SVZ}, 
a few thousand papers have been published
on applications of the QCD sum rule method in all corners of low energy
hadronic physics. Unavoidably, results from different collaborations 
were not always consistent \cite{REVIEW}.
The main reason for these
inconsistencies was frequently the impossibility of  estimating reliably the
errors in the method. With the advent of precise measurements of the vector
(V) and axial-vector (A) spectral functions, obtained from tau-lepton
decay \cite{ALEPH}-\cite{OPAL}, an opportunity was opened to check the
precision of the QCD sum rules in the light-quark sector of QCD. In this note
we would like to present a critical and conservative appraisal of chiral sum
rules of the Weinberg type \cite{WSR}, as they are
confronted with experimental data for the  spectral functions. This kind
of sum rules involve the difference between the vector and the axial-vector
correlators (V-A), which vanishes identically to all orders in  perturbative
QCD in the chiral limit. In fact, neglecting the light quark masses, the
(V-A) two-point function vanishes like  $1/q^{6}$ in the space-like region,
where the scale ${\cal {O}}(300$ MeV) is set by the quark and gluon
condensates. In the time-like region the chiral spectral
function $\rho _{V-A}(q^{2})$ should also vanish for large
$Q^2\equiv - q^{2}$, but judging from the ALEPH data\cite{ALEPH}, the
asymptotic regime of local
duality may not have been reached in $\tau $-decay . Under less stringent 
assumptions one expects global duality to hold in the time like region;
in particular, this should be the
case for the Weinberg-type sum rules. Surprisingly, these sum rules also
appear to be poorly convergent.  A possible source of duality
violation could be some non-perturbative contribution to the correlator (e.g.
due to instantons) which falls off exponentially in the space-like region but
oscillates in the time-like region. If the duality violations were due to this
source, then there would be a simple recipe (introduced 30 years ago \cite
{Nasrallah}) to improve convergence. \\
In a previous publication \cite{DS1} we studied some QCD chiral sum rules of
the Weinberg type, and their saturation by the ALEPH data. In
particular, we showed that a remarkable improvement of this saturation can
be achieved by introducing a polynomial integration kernel which vanishes
at the upper limit of integration. However, no detailed quantitative error
analysis was performed in \cite{DS1}. In this note we reexamine the
saturation of several QCD chiral sum rules using the ALEPH \cite{ALEPH}, as
well as the OPAL data \cite{OPAL}, and paying particular attention to the
error analysis. We obtain an updated determination of $\bar{L}_{10}$, the
scale independent part of the coupling constant of the relevant operator in
the ${\cal {O}}$$(p^{4})$ counter terms in the Lagrangian of chiral
perturbation theory \cite{GL}. This quantity is related to the finite
remainder of the (V-A) correlator at zero momentum. We also determine the
finite remainder of the first derivative of the (V-A) correlator at zero
momentum, which is related to the ${\cal {O}}$$(p^{6})$ counter terms.
Finally, we introduce combinations of QCD chiral sum rules which allow for a
determination of the (V-A) dimension $d=6$ and  $d=8$
vacuum condensates. The former can be extracted with reasonable precision,
while the latter is affected by much larger uncertainties. \\

We begin by defining the vector and axial-vector current correlators
\begin{eqnarray*}
\Pi_{\mu \nu}^{VV} (q^2) = i \; \int \; d^4 \; x \; e^{i q x} \; \;
<0|T(V_{\mu}(x) \; \; V_{\nu}^{\dagger}(0))|0> \;
\end{eqnarray*}
\begin{equation}
= \; (- g_{\mu \nu} \; q^{2} + q_{\mu} q_{\nu}) \; \Pi_{V} (q^{2}) \; ,
\end{equation}
\begin{eqnarray*}
\Pi_{\mu \nu}^{AA} (q^2) = i \; \int \; d^4 \; x \; e^{i q x} \; \;
<0|T(A_{\mu}(x) \; \; A_{\nu}^{\dagger} (0) )|0> \;
\end{eqnarray*}
\begin{equation}
= \; (- g_{\mu \nu} \; q^{2} + q_{\mu} q_{\nu}) \; \Pi_{A} (q^{2}) - q_{\mu}
q_{\nu} \; \Pi_{0} (q^{2}) \; ,
\end{equation}

where $V_{\mu }(x)=:\bar{q}(x)\gamma _{\mu }q(x):$, $A_{\mu }(x)=:\bar{q}
(x)\gamma _{\mu }\gamma _{5}q(x):$, and $q=(u,d)$. Here we shall concentrate
on the chiral correlator $\Pi _{V-A}\equiv \Pi _{V}-\Pi _{A}$. This
correlator vanishes identically in the chiral limit ($m_{q}=0$), to all
orders in QCD perturbation theory. Renormalon ambiguities are thus avoided.
Non-perturbative contributions due to vacuum condensates contribute to this
two-point function starting with dimension $d=6$ and involving the four-quark
condensate. The Operator Product Expansion (OPE) of the chiral correlator
can be written as

\begin{equation}
\Pi(Q^2)|_{V-A} = \sum_{N=1}^{\infty} \frac{1}{Q^{2N+4}}\;
C_{2N+4}(Q^2,\mu^2)\; <{\cal {O}}_{2N+4}(\mu^2)> \;,
\end{equation}

with $Q^{2}\equiv -q^{2}$. It is valid away from the positive real axis for
complex $q^{2}$, and $\left\vert q^{2}\right\vert $ large. Radiative
corrections to the $d=6$ contribution are known \cite{CH}. They
depend on the regularization scheme, implying that the value of the
condensate itself is a scheme dependent quantity. Explicitly,

\begin{equation}
\Pi(Q^2)|_{V-A}\;=\;- \frac{32\pi }{9}\;\frac{\alpha _{s}<\bar{q}q>^{2}}{
Q^{6}}\;\left\{ 1+\frac{\alpha _{s}(Q^{2})}{4\pi }\;\left[ \frac{247}{12}+
{\rm ln}\left( \frac{\mu ^{2}}{Q^{2}}\right) \right] \right\} \;+{\cal O}
(1/Q^{8})\;,
\end{equation}

in the anti-commuting $\gamma _{5}$ scheme, and assuming vacuum saturation
of the four-quark condensate. Radiative corrections for  $d \geq
8$ are not known.
To facilitate comparison with current conventions in the literature it will
be convenient to absorb the Wilson coefficients, including radiative
corrections, into the operators, and rewrite Eq.(3) as

\begin{equation}
\Pi(Q^2) = \sum_{N=1}^{\infty} \frac{1}{Q^{2N+4}} <{\cal {O}}_{2N+4}> \;,
\end{equation}

where we have dropped the subscript (V-A) for simplicity.
We will be concerned with Finite Energy Sum Rules of the type

\begin{equation}
W(s_0) \equiv \int_{0}^{s_0} \; ds \; f(s) \rho (s) \; ,
\end{equation}

where $f(s)$ is a weight function, and the hadronic spectral function $\rho
(s)\equiv \rho _{V}(s)-\rho _{A}(s)$, with $\rho _{V,A}(s)=\frac{1}{\pi }
Im\Pi _{V,A}(s)$ (pion pole excluded from $\rho _{A}(s)$).
For instance, if $f(s)=s^{N}$ ($N=0,1,2,...$), then one obtains

\begin{equation}
\int_{0}^{s_0} \; ds \; s^N \rho (s) = f_\pi^2 \;\delta_{N0} + (-)^N <{\cal {
O}}_{2N+2}> \;\; (N=0,1,2,...) \; ,
\end{equation}

where $f_{\pi} = 92.4 \pm 0.26 \; \mbox{MeV}$ \cite{PDG}. For $N=0,1$ Eq.(8)
leads to the first two (Finite Energy) Weinberg sum rules, while for $N=2,3$
the sum rules project the $d=6,8$ vacuum condensates, respectively; notice
that in the chiral limit $<{\cal{O}}_2> = <{\cal{O}}_4> = 0$. To
first order in $\alpha_s$, radiative corrections to the vacuum condensates
do not induce mixing of condensates of different dimension in a given FESR 
\cite{MIX}. We shall also consider the chiral correlator, and its first
derivative, at zero momentum; the finite remainder of these being given by
the sum rules

\begin{equation}
\bar{\Pi}(0) = \int_{0}^{s_0} \; \frac{ds} {s}\; \rho(s) \; ,
\end{equation}

\begin{equation}
\bar{\Pi}^{\prime}(0) = \int_{0}^{s_0} \; \frac{ds} {s^2}\; \rho(s) \; ,
\end{equation}

where $\rho (s)$ does not contain the pion pole. Equation (8) is the
Das-Mathur-Okubo (Finite Energy) sum rule \cite{WSR} . The finite remainder $
\bar{\Pi}(0)=-4\bar{L}_{10}$ ,where $\bar{L}_{10}$ is a counter term of the $
{\cal {O}}$$(p^{4})$ Lagrangian of chiral perturbation theory,
can be expressed as

\begin{equation}
\bar{\Pi}(0)=-4\bar{L}_{10}=\left[ \frac{1}{3}f_{\pi }^{2}<r_{\pi
}^{2}>-F_{A}\right] \;  = 0.026 \pm 0.001 \;,
\end{equation}

where $<r_{\pi }^{2}>$ is the electromagnetic mean squared radius of the
pion, $<r_{\pi }^{2}>=0.439\pm 0.008\;\mbox{fm}^{2}$ \cite{AMEN}, and $F_{A}$
is the axial-vector coupling measured in radiative pion decay, $
F_{A}=0.0058\pm 0.0008$ \cite{PDG}.
Similarly, $\bar{\Pi}^{\prime }(0)$ is
related to the ${\cal {O}}$$(p^{6})$ counter terms.

As mentioned earlier, the saturation of the various chiral sum rules can be
considerably improved by introducing an integration kernel that vanishes at
the upper limit of integration ($s=s_{0}$). We have tested a variety of such
kernels searching for optimal saturation. The following results have been
obtained using the ALEPH data for $\rho (s)$, with the errors at each energy
bin calculated from the error correlation matrix. Use of the OPAL data \cite
{OPAL} data leads to similar results, albeit with much lager error
bands. Starting with the first Weinberg sum rule, Fig.1 shows the left hand
side of Eq.(7) for $N=0$ (curve(a)), together with the right hand side, i.e.
$f_{\pi }^{2}$ (straight line (c)), as well as the modified sum rule
(curve(b))

\begin{equation}
W_1(s_0) = \int_{0}^{s_0} \; ds \; (1-\frac{s}{s_0} ) \; \rho(s) \; .
\end{equation}

On account of the second Weinberg sum rule, curves (a) and (b) should be
identical; the improved saturation achieved with Eq.(11) being remarkable.
Figure 1 can be used to present our criterion to judge the reliability of a
QCD sum rule. The sum rule must be presented explicitly as a function of
the upper integration limit $s_{0}$. If the left hand side is a constant,
then the spectral integral must also be  approximately a constant,
starting from 1 to 2 GeV$^{2}$ up to the maximum s$_{0}$ of the data.
From Fig.1 we would extract
\begin{equation}
f_{\pi }^{2} =0.008\pm 0.004 \;\mbox{GeV}^2\; ,
\end{equation}
for curve (a), and
\begin{equation}
f_{\pi }^{2} =0.0084\pm 0.0004 \;\mbox{GeV}^2\;,
\end{equation}
for curve (b), to be compared with the experimental value
$f_{\pi }^{2}|_{EXP} =0.00854\pm 0.00005 \;\mbox{GeV}^2\;.$
Curve (a) demonstrates the fact that if the spectral integral is not a
constant then the experimental errors are quite irrelevant in a test of
duality. It is very dangerous to pick up a small  stability region
to obtain a prediction (here one could choose the region around 2 GeV$^{2}$).

In Fig.2 we show Eq.(8) (curve(a)) together with the modified sum rule
(curve (b))

\begin{equation}
\bar{\Pi}(0) = 2 \;\frac{f_\pi^2}{s_0} \; + \;\int_{0}^{s_0} \; \frac{ds} {s}
\; (1-\frac {s}{s_0})^2 \;\rho(s) \; .
\end{equation}

From the optimized sum rule (14) we obtain the value (straight line (c)) 
\begin{equation}
\bar{\Pi}(0)=-4\bar{L}_{10}=0.0257\pm 0.0003 \;,
\end{equation}
which is considerably more accurate than the leading order chiral
perturbation
theory result, Eq. (10). The agreement between Eqs. (10) and (15)
may be an indication that  higher order chiral corrections to the
Das-Mathur-Okubo sum rule are indeed very small.
Figure 3 shows Eq.(9) (curve (a)) together with
the optimized sum rule (curve(b))

\begin{equation}
\bar{\Pi}^{\prime}(0) = \frac{3}{s_0}\; \bar{\Pi}(0) - 3 \; \frac{f_\pi^2}{
s_0^2} + \int_{0}^{s_0} \; \frac{ds} {s^2}\; (1-\frac{s}{s_0} )^3 \;\rho(s)
 \;
,
\end{equation}

the latter giving (curve(c))
\begin{equation}
\bar{\Pi}^{\prime}(0) = 0.065 \pm 0.001 \;\mbox{GeV}^{-2} \;.
\end{equation}
We turn now to the determination of the $d=6$ and $d=8$ vacuum condensates.
In Fig. 4 we show $<{\cal {O}}_{6}>$ as obtained from Eq.(7) with $N=2$
 (curve
(a)), together with the result from the improved sum rule (curve(b))

\begin{equation}
<{\cal {O}}_{6}> = - f_\pi^2\; s_0^2 + s_0^2 \int_{0}^{s_0} \; ds\;
 (1 -\frac{s}{s_0} )^2
\;\rho(s) \; ,
\end{equation}
which gives
\begin{equation}
<{\cal {O}}_{6}> = - (0.004 \pm 0.001) \;\mbox{GeV}^6 \;.
\end{equation}
This result can be compared with the vacuum saturation expression
\begin{equation}
<{\cal {O}}_{6}>|_{VS} =- \frac{32}{9} \;\pi\; \bar{\alpha_s} |<\bar{q} q>|^2
 \;
\simeq  - 1.1 \times 10^{-3} \;\mbox{GeV}^6 \;,
\end{equation}
to leading order
in $\alpha_s$, and
where we used $<\bar{q} q> = - 0.014 \;\mbox{GeV}^3$,
and $\bar{\alpha_s} = 0.5$,
at  a scale of 1 GeV. Radiative corrections increase this estimate by a
factor of two.
The result Eq. (19) confirms pioneer determinations from
$e^+ e^-$, as well as tau-lepton decay data \cite{BERTL}-\cite{DSO}
indicating
that the vacuum saturation approximation underestimates the $ d = 6$
condensate
roughly by a factor of 2-3.

Finally, for $<{\cal {O}}_{8}>$ Fig.5 (curve(a)) shows the result from
Eq.(7) with $N=3$, together with the improved determination from the sum rule
(curve(b))

\begin{equation}
<{\cal {O}}_{8}> = 8\; s_0^3\; f_\pi^2 - 3 \;s_0^4\; \bar{\Pi}(0) +
s_0^3 \int_{0}^{s_0} \; \frac{ds}{s} \; (1 - \frac{s}{s_0} )^3 \;(s + 3 s_0)
 \; \rho(s) \; ,
\end{equation}

which gives
\begin{equation}
<{\cal {O}}_{8}> = - (0.001 \pm 0.006) \;\mbox{GeV}^8 \;,
\end{equation}
in the region where the condensate is approximately constant 
($ s_0 \simeq 1.75 - 2.5 \; \mbox{GeV}^2$), and
assuming, optimistically, that the stability region has
been reached beyond $s_0 \simeq 2.5\; \mbox{GeV} ^2$.
It should be clear from Fig. 5 that no meaningful determination of
$<{\cal {O}}_{8}>$ is possible using the standard FESR, Eq.(7).
As expected, with increasing dimensionality, i.e higher powers of $s$ in the
dispersive integrals, the accuracy of the determination of the vacuum
condensates deteriorates considerably. It should be noticed that the results
(19) and (22) do not rely on the vacuum saturation approximation. They also
include all radiative corrections, and are correct to first order in $
\alpha_s$. At order $\alpha_s^2$ and beyond, there is no longer
decoupling of condensates of different dimensionality in a given FESR \cite
{MIX}. However, one expects these higher order radiative corrections to the
Wilson coefficients in the OPE to be small.
There seems to be general agreement in the literature
on the size of the $d=6$ condensate, but there exists a number of
inconsistent QCD sum rule determinations of the value of the $d=8$
condensate. The results range from $<O_8> = -(3.5\pm 2.0)\times 10^{-3}$
 GeV$^{8}$ \cite{IZ} to $<O_8> = (4.4\pm 1.2) \times 10^{-3}$ GeV$^{8}$
 \cite{DGHS}. Our result is consistent, within the large errors, with
a recent determination \cite{O8} (this reference contains a detailed
comparative study of the literature).\\

The poor convergence of ordinary QCD chiral sum rules is rather intriguing,
as one would have expected good saturation at relatively low energies, given
the very rapid fall-off of the chiral (V-A) correlator (see Eq.(4)).
However, extrapolating the chiral correlator from the space-like to the
time-like region can produce strong changes close to the real-axis. In fact,
violations of local duality at the 100\% level have been shown to be
possible using realistic models of the heavy quark chiral correlator \cite
{BARI}. The remarkable improved saturation achieved by introducing weight
functions that vanish on the real axis at $s=s_{0}$ could be taken as an
indication that although perturbative QCD works well in the space-like
region, this may not be the case in the time-like region, or  near the cut,
at least at
energies below $s_{0}\simeq 3.5\;\mbox {GeV}^{2}$. Finally, by using the
chiral (V-A) correlator we have been able to extract the value of the $d=6$
vacuum condensate with reasonable accuracy; for the $d=8$ condensate the
result is affected by a large uncertainty. In contrast, were one
to attempt a determination from the vector correlator, and separately from
the axial-vector one, the results would be quite inconclusive. This is due
to the very large current value of $\Lambda _{QCD}$ ($\Lambda _{QCD}\simeq
400\;\mbox{MeV})$ which makes the perturbative QCD term in the OPE so big
that it overwhelms the power corrections. We estimate that if $\Lambda
_{QCD}\simgt\; 330\;\mbox{MeV}$ then the FESR, and even Laplace transform sum
rules, will be unable to provide a conclusive determination of the vacuum
condensates. Earlier standard extractions of these condensates from
electron-positron annihilation \cite{BERTL} and tau-lepton decay \cite{DSO}
relied on past values $\Lambda _{QCD}\simeq 100-200\;\mbox{MeV}$. With these
values of $\Lambda _{QCD}$ the perturbative QCD term in the OPE is dominant
but not overwhelming, and the power corrections can be clearly discerned.\\

We would like to conclude with a general comment. Mathematically, the
extraction of QCD parameters from experiment via sum rules constitutes a so
called ill posed inverse problem (analytic continuation of an imprecisely
known function). Small changes in the input data lead to large changes in
the output. The problem is stabilized by extracting only a small number of
parameters. Given the present accuracy of the $\tau $-decay data, we
conclude from our analysis that only the condensate\ $<{\cal {O}}_{6}>$ can
be extracted with some degree of confidence, and only a rough idea of the
order
of magnitude of $\ <{\cal {O}}_{8}>$can be obtained. This situation cannot
be remedied by mathematical tricks like employing Laplace or Gaussian
integration kernels.  Only with forthcoming more accurate data, can one
expect to extract higher dimensional condensates.

\newpage
\begin{center}
{\bf Figure Captions}
\end{center}

Figure 1. Curve (a) is the standard first Weinberg sum rule, Eq.(7)
with $N=0$, curve (b) is the modified sum rule Eq. (11), and curve (c)
is the experimental value of $f_\pi^2$. \newline

Figure 2. The chiral correlator at zero momentum, $\bar{\Pi}(0)$,
from the standard sum rule Eq. (8) (curve (a)), and from the modified
sum rule Eq.(14) (curve (b)), the latter leading to the prediction Eq. (15)
 (curve(c)).
\newline

Figure 3. The first derivative of the chiral correlator at zero momentum,
$\bar{\Pi}'(0)$, from the standard sum rule Eq.(9) (curve (a)), and from the
modified sum rule Eq.(16), the latter leading to the prediction Eq.(17)
 (curve (c)).
\newline

Figure 4. The dimension-six vacuum condensate from the standard sum rule,
Eq. (7) with $N=2$ (curve (a)), and from the modified sum rule Eq.(18)
(curve (b)).\newline

Figure 5. The dimension-eight vacuum condensate from the standard sum rule,
Eq. (7) with $N=3$ (curve (a)), and from the modified sum rule Eq.(21)
(curve (b)).

\newpage

\begin{figure}[tp]
\epsffile{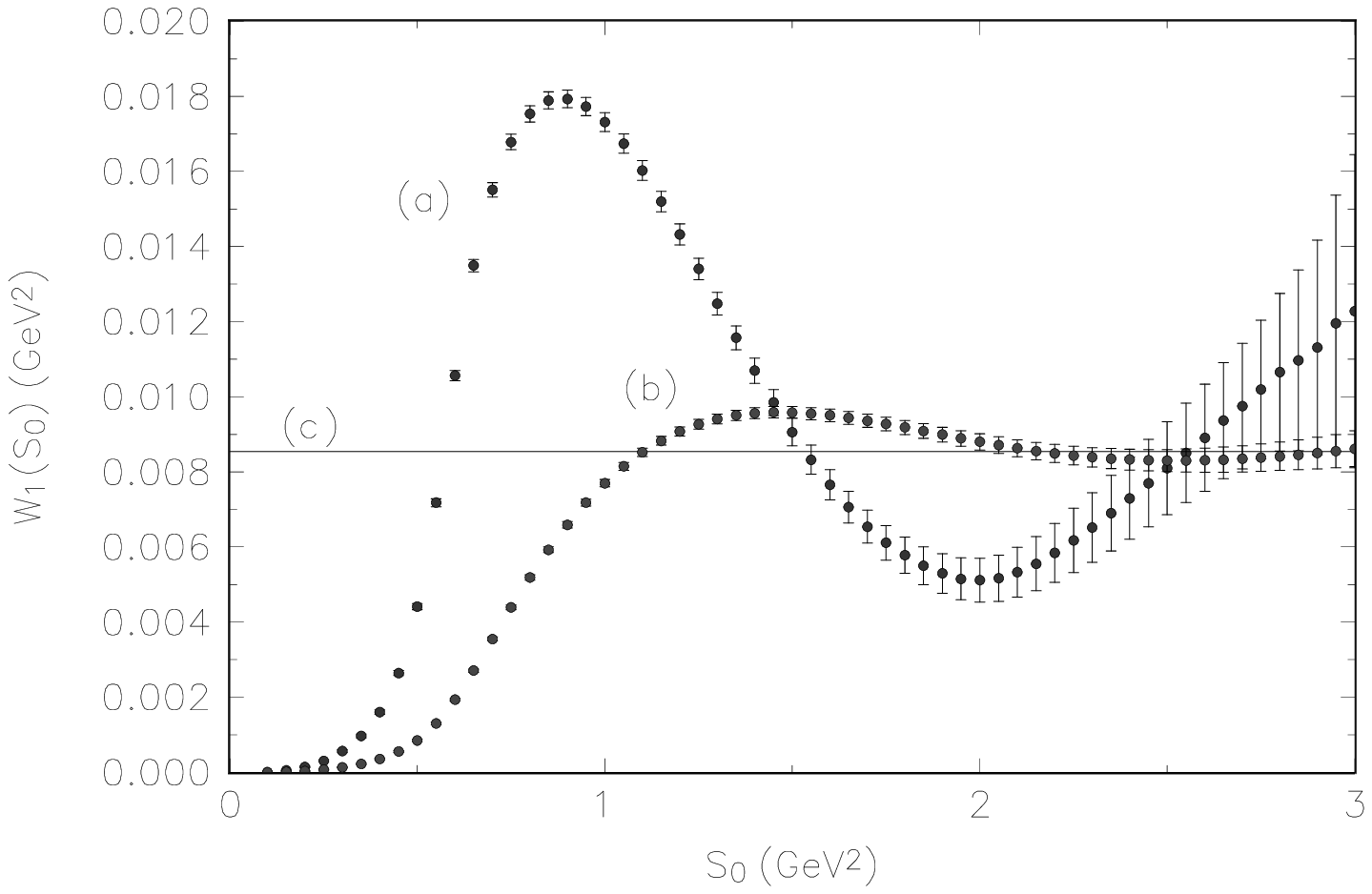}
\caption{}
\end{figure}
\newpage
\begin{figure}[tp]
\epsffile{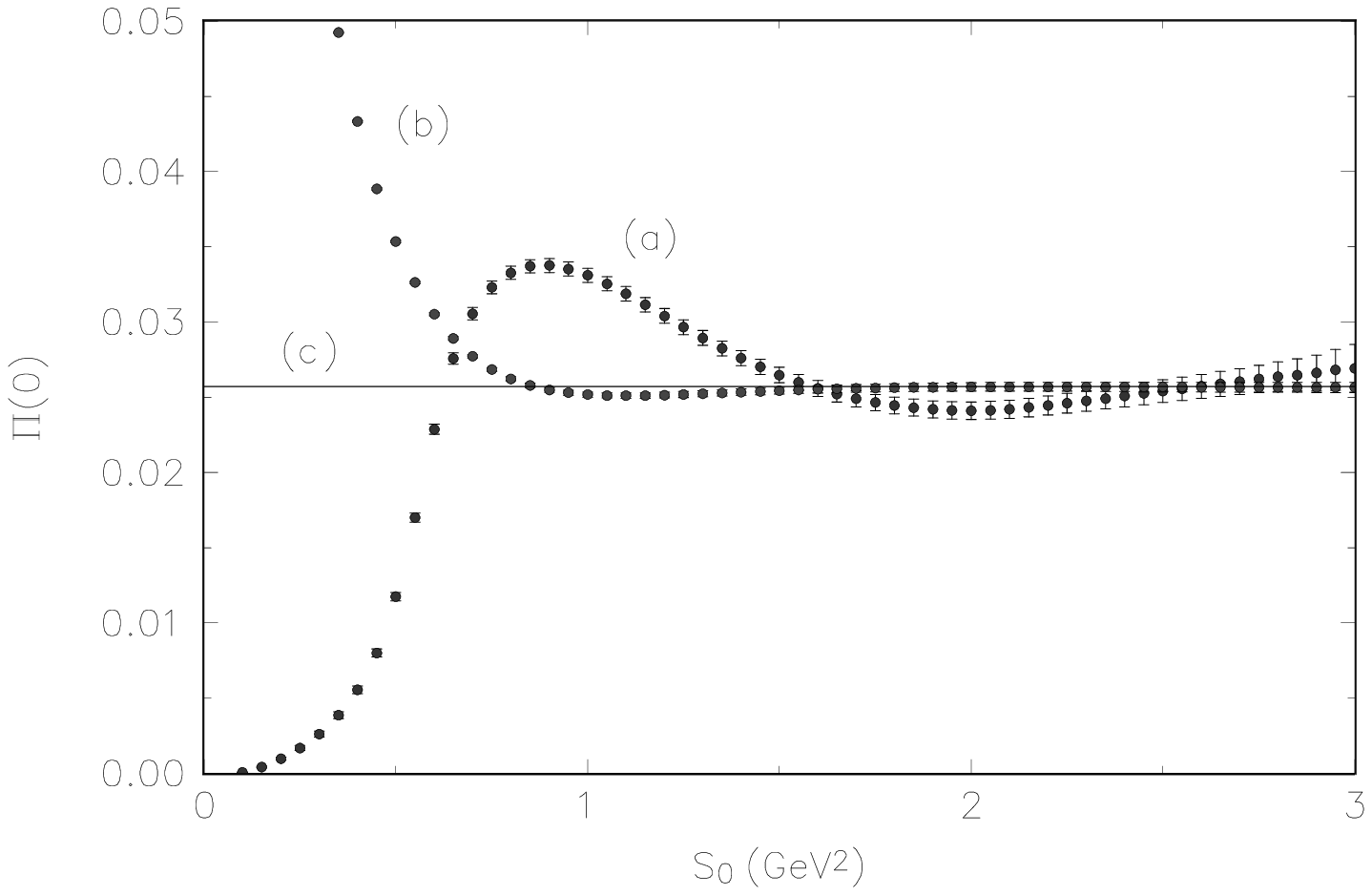}
\caption{}
\end{figure}
\newpage
\begin{figure}[tp]
\epsffile{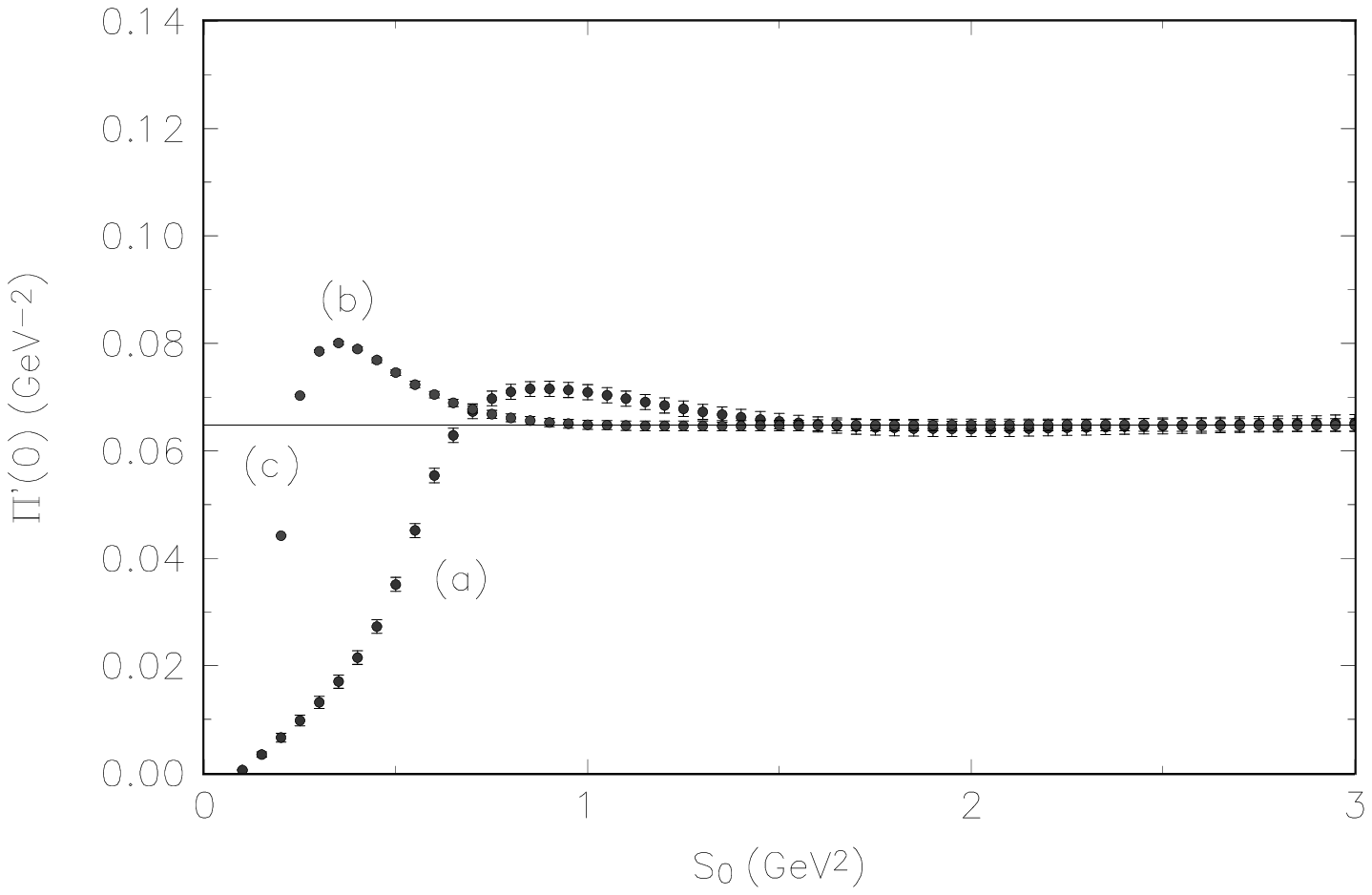}
\caption{}
\end{figure}
\newpage
\begin{figure}[tp]
\epsffile{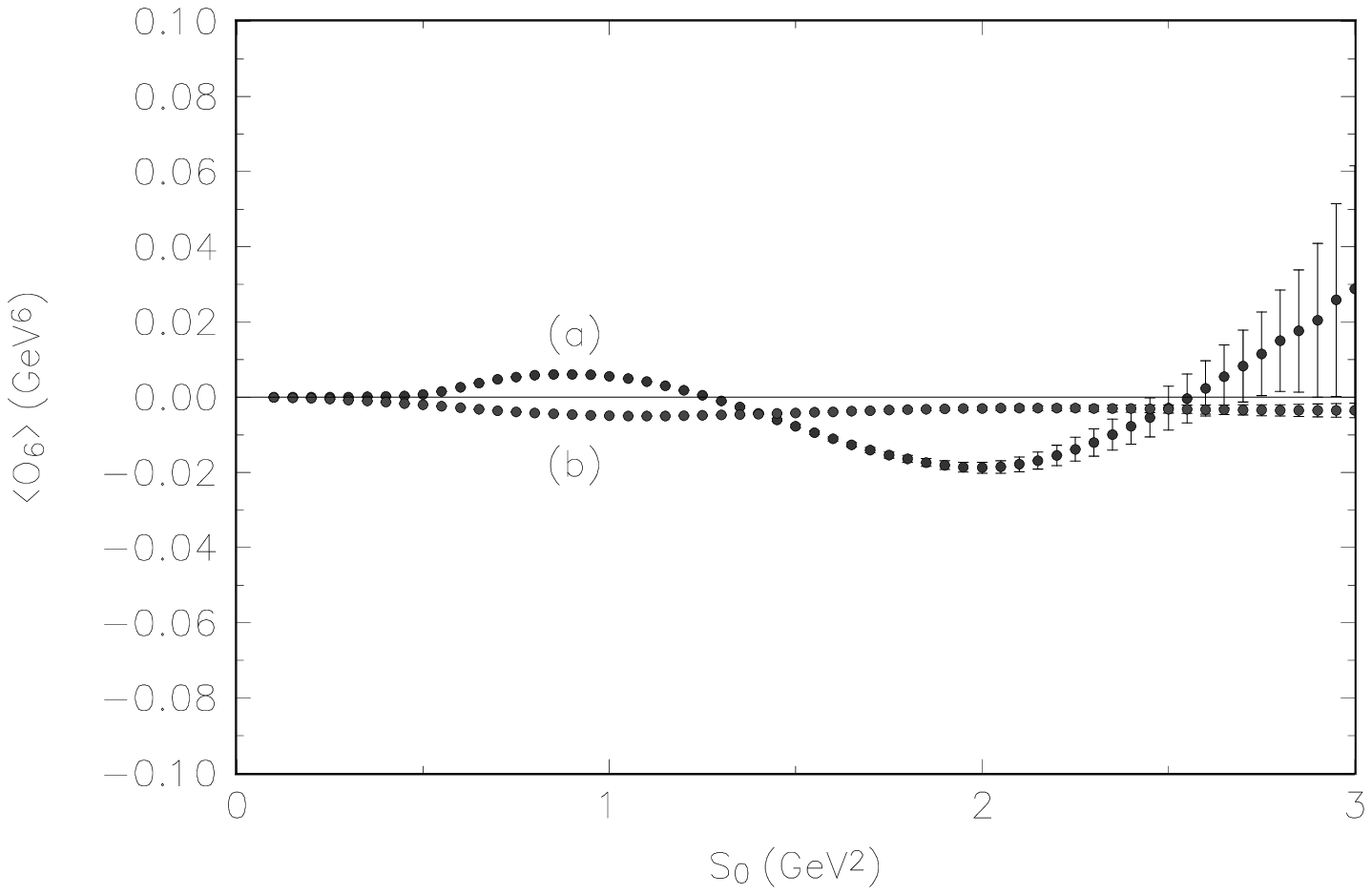}
\caption{}
\end{figure}
\newpage
\begin{figure}[tp]
\epsffile{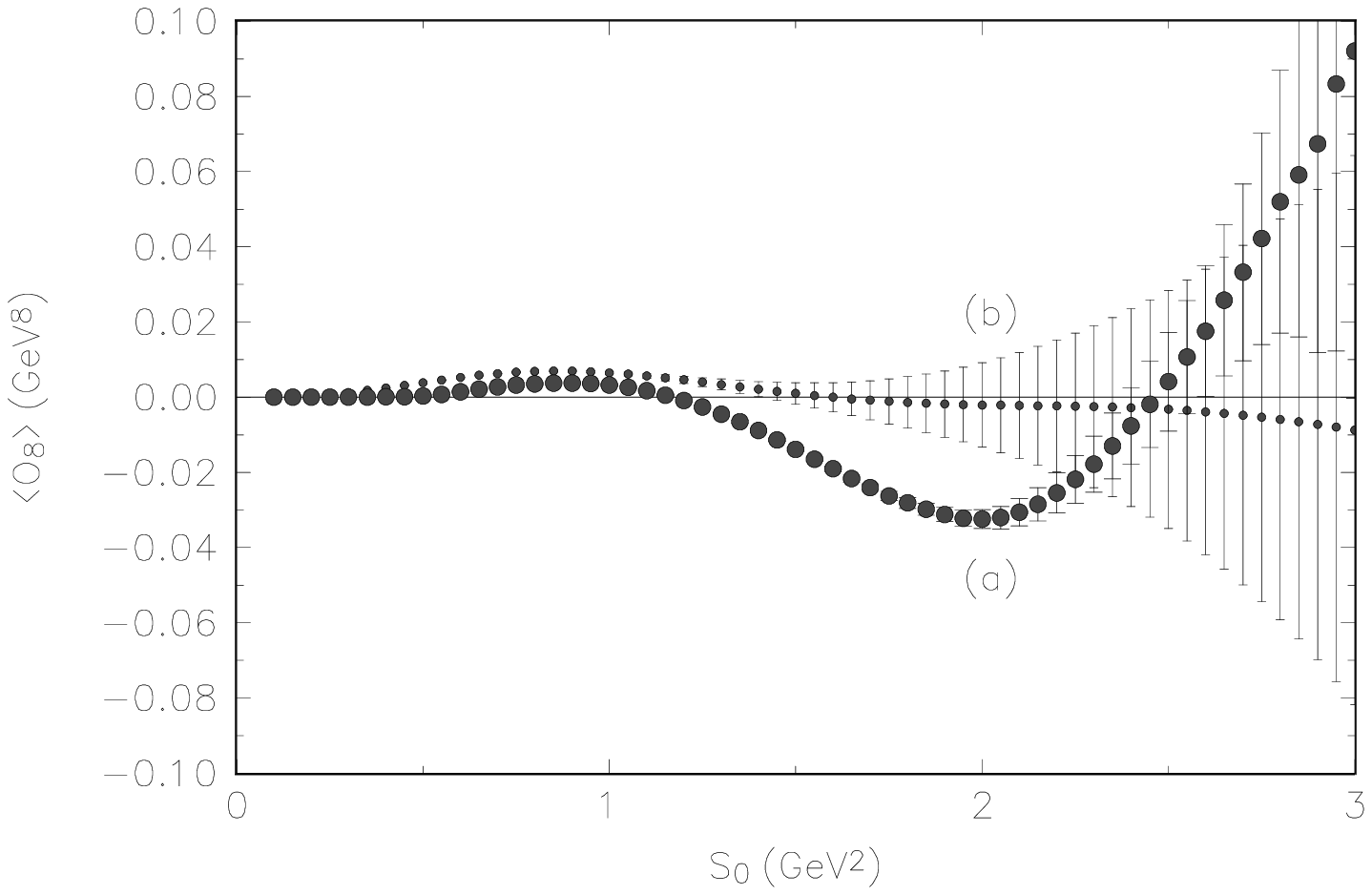}
\caption{}
\end{figure}

\end{document}